# Investigation of the Short Argon Arc with Hot Anode, Part I: Numerical Simulations of Non-equilibrium Effects in the Near-electrode Regions

A. Khrabry[1], I.D. Kaganovich[1], V. Nemchinsky[2], A. Khodak[1]

[1]*Princeton Plasma Physics Laboratory, Princeton NJ, 08543 USA*

[2]*Keiser University, Fort Lauderdale, FL, 33309 USA*

## Abstract

Atmospheric pressure arcs have recently found application in the production of nanoparticles. Distinguishing features of such arcs are small length and hot ablating anode characterized by intensive electron emission and radiation from its surface. We performed one-dimensional modeling of argon arc, which shows that near-electrode effects of thermal and ionization non-equilibrium play important role in operation of a short arc, because the non-equilibrium regions are up to several millimeters long and are comparable with the arc length. The near-anode region is typically longer than the near-cathode region and its length depends more strongly on the current density. The model was extensively verified and validated against previous simulation results and experimental data. Volt-Ampere characteristic (VAC) of the near-anode region depends on the anode cooling mechanism. The anode voltage is negative. In case of strong anode cooling (water-cooled anode) when anode is cold, temperature and plasma density gradients increase with current density resulting in decrease of the anode voltage (absolute value increases). Falling VAC of the near-anode region suggests the arc constriction near the anode. Without anode cooling, the anode temperature increases significantly with current density, leading to drastic increase in the thermionic emission current from the anode. Correspondingly, the anode voltage increases to suppress the emission – and the opposite trend in the VAC is observed. The results of simulations were found to be independent of sheath model used: collisional (fluid) or collisionless model gave the same plasma profiles for both near-anode and near-cathode regions.

## I. Introduction

Atmospheric pressure arcs recently found application in production of nanoparticles, such as carbon nanotubes[1,2,3,4,5] and boron-nitride nanotubes[6]. Distinguishing features of such arcs are typically short length (about several millimeters inter-electrode gap) and hot ablating anode, which provides feedstock material for growth of nanoparticles. High anode temperature leads to thermionic electron emission and intensive radiation from its surface. The high electron emission from the hot anode strongly affects the near-anode plasma, especially the space-charge sheath and the heat flux to the anode. Because the arc is short, non-uniform near-electrode regions of thermal and ionization non-equilibrium play important role in the arc physics.

Electrode ablation significantly increases complexity of the arc physics, because such arc plasma consists of mixture of gases with different ionization and transport characteristics. These processes will be considered in follow-up publications. In this paper we only focus on non-equilibrium processes in the arc with non-ablating electrodes, e.g., a short argon arc with tungsten electrodes.

Though arcs has been extensively studied earlier, most of the studies considered only specific aspects of the arc physics: cathodic region[7,8,9,10,11,12,13], arc column[14,15], anodic region[16,17,18,19,20,21,22,23,24]. A detailed review of works on argon arc modeling can be found in Ref. 25. However, we could not find papers considering arc with hot emitting anode. Thorough reviews on numerical and experimental studies of the near-anode region of arc discharges can be found in Refs. 26 and 27. Typically in simulations of the whole arc, thermal or chemical equilibrium plasma model (Saha equation) is used[28,29,30,31,32] or space-charge sheath effects are not taken into account[33,34,35]. Effects of near-electrode non-equilibrium layers were taken into account in recent arc simulations[36] making use of specific boundary conditions; bulk plasma was considered equilibrium. The layers were considered infinitely thin, which may be inaccurate for short arcs. Numerical simulation of the whole arc making use of the non-equilibrium plasma model was only recently reported in Ref. 37. The authors of Refs. 36 and 37 focused primary on plasma-cathode interaction and for the anode some simplifications were still used. A thorough non-equilibrium fluid model of plasma of the atmospheric and high-pressure arcs was developed in Ref. 13. Governing equations for species transport and heat transfer, and coefficients therein are derived from kinetic theory[38,39]. The model is validated by comparison with experimental data, in particular, see validation for cathodic part of argon arc in Ref. 13.

In the present study, this model[13] was expanded and implemented into one-dimensional (1D) code for self-consistent simulations of the whole arc, including heat transfer in cylindrical electrodes, and radiation from their surfaces. For short arcs, in which axial gradients are much higher than radial, 1D approximation is rather accurate. The code results were benchmarked against previous simulations[13] and validated against experimental data[40,41]. Different models of space-charge sheath regions were implemented into the code: solving the Poisson equation for the collisional sheath-plasma[13] and the quasineutral plasma with sheath boundary conditions for the collisionless sheath model[42,43]. The models were applied to both near-anode and near-cathode regions and compared to each other.

Parametric studies of short atmospheric pressure argon arc with tungsten electrodes were performed for various current densities and inter-electrode gap sizes. Non-equilibrium effects in the near-electrode regions were studied. Anodes with and without water cooling were considered. Effect of electron emission on current-voltage characteristic of the near-anode layer was investigated. Analytical formulas for scaling of non-equilibrium regions widths and Volt-Ampere's characteristics of these regions and the whole arc are given in the accompanying paper[44].

The organization of the paper is as follows. In Section II governing equations and boundary conditions for plasma and electrodes are presented. Section III describes numerical procedure of solving the governing equations. Results of simulations including validation of the model and parametric studies of the arc are presented and discussed in Section IV. Conclusions of this work are summarized in Section V.

## II. The 1D arc model

The model consists of equations for the plasma region, corresponding boundary conditions and equations for heat transport in the electrodes. The equations for the plasma region are derived from the kinetic theory, see Ref. 13 and references therein for details.

### II.1. Governing equations for plasma region

**Momentum balance of species:**

Momentum balance of individual species in multi-component mixture can be described by Stefan-Maxwell equations[39], which accurately account for the species cross-diffusion:

$$-\nabla p_k + \frac{\rho_k}{\rho}\left(\nabla p - e(n_i - n_e)\vec{E}\right) + n_k e Z_k \vec{E} - \sum_{j \neq k} \frac{n_k n_j k T_{kj} C_{kj}}{n D_{kj}}(\vec{v}_k - \vec{v}_j) - \vec{R}_k^{th} = 0, \quad (1)$$

where $k, j$ are subscripts, denoting different species: argon atoms $a$, argon ions $i$ and electrons $e$; $n_k$, $n$ are the number density of species $k$ and plasma and background gas as a whole; $\rho_k$, $\rho$ are the mass density of species $k$ and plasma and background gas as a whole; $\vec{v}$ is the mean (mass-averaged) velocity; $\vec{E}$ is the electric field; $k$ is the Boltzmann constant; $e$ is the elementary charge; $Z_k$ is the charge number: -1 for electrons, 1 for ions, 0 for neutrals; $C_{kj}$ are the coefficients derived from the kinetic theory (typically about unity); $C_{ia} = 1$, $C_{ei} = 0.506$, definitions of other coefficients are more complicated and can be found in Ref. 13. $\vec{R}_k^{th} = C_k^{(e)} n_k k \nabla T_e$ is the thermal-diffusion force, where the term with gradient of temperature of heavy particles is neglected. Kinetic coefficients $C_k^{(e)}$ are defined as follows:

$$C_i^{(e)} = -\frac{n_e}{n_i}\frac{1}{1+P}C_{tdei}, \quad C_a^{(e)} = -\frac{n_e}{n_a}\frac{P}{1+P}C_{tdea}, \quad C_e^{(e)} = \frac{1}{1+P}C_{tdei} + \frac{P}{1+P}C_{tdea},$$

$P$ is the ratio of electron-atom and electron-ion collision frequencies:

$$P = \frac{\nu_{e,a}}{\nu_{e,i}} = \frac{n_a D_{ei}}{n_i D_{ea}}, \quad (2)$$

$C_{tdei}$, $C_{tdea}$ are the values of thermal diffusion coefficient for electrons in the limits of strongly and partially ionized plasmas. $C_{tdei} = 0.703$ and $C_{tdea}$ is defined via integral of electron-atom collision cross-section[13,38]. $D_{kj}, T_{kj}$ are the binary diffusion coefficients and temperatures defined as follows:

$$nD_{kj} = \frac{3\pi}{32}\sqrt{\frac{8kT_{kj}}{\pi m_{kj}}}\frac{1}{\sigma_{kj}}, \quad m_{kj} = \frac{m_k m_j}{m_k + m_j}, \quad T_{kj} = \frac{m_k T_j + m_j T_k}{m_k + m_j}, \quad (3)$$

$m_k$, $T_k$ are the mass of particles of sort k, is their temperature. Temperatures and masses of heavy particles are very close and are not distinguished in the model: $T_i = T_a = T$, $m_i = m_a = m_{Ar}$. $\sigma_{kj}$ are the momentum transfer cross sections. For electron-atom collisions, cross-section data was taken from Ref. 45, approximate value of electron-atom momentum transfer cross-section is $3 \cdot 10^{-20} m^2$. For ion-atom interactions charge-exchange cross-section was used[46]: $\sigma_{ia} = 9.2 \cdot 10^{-19} m^2$. Coulomb collisions between electrons and ions can be described by the following cross-section[47]:

$$\sigma_{ei} = \frac{e^4 \ln \Lambda}{32\pi\varepsilon_0^2 (kT_e)^2}, \qquad (4)$$

where $\varepsilon_0$ is the vacuum permittivity, and $\ln \Lambda = \ln\left(8\pi\varepsilon_0 kT_e \sqrt{\varepsilon_0 kT_e / n_e} / e^3\right)$ is the Coulomb logarithm.

Note that equation (1) is written in a general multi-dimensional form. In the code, it was implemented in 1D form. In 1D approximation for non-ablating electrodes the mean velocity is zero everywhere and momentum conservation equation for the mixture as a whole takes the form:

$$\nabla p - e(n_i - n_e)\vec{E} = 0. \qquad (5)$$

Substitution of relation (5) into Eq. (1) eliminates the second term in (1).

### Species conservation equations:

For each of the species, the conservation equation can be written in the following form:

$$\nabla \Gamma_k = s_k, \qquad (6)$$

where: $\vec{\Gamma}_k = n_k \vec{v}_k$ is the flux of species k, $s_k$ is the volumetric source:

$$-s_a = s_i = s_e = k_i n_e n_a - k_r n_i n_e^2, \qquad (7)$$

where $k_i$, $k_r$ are the reaction rate coefficients. $k_i$ is calculated using approach described in Ref. 48, $k_r$ is calculated making use of the ionization equilibrium condition:

$$k_i n_{e,Saha} n_a - k_r n_i n_{e,Saha}^2 = 0, \qquad (8)$$

where $n_{e,Saha}$ is the equilibrium number density defined by the Saha equation:

$$\frac{n_{e,Saha}^2}{n_a} = 2g_i / g_a \left(\frac{2\pi m_e kT_e}{h^2}\right)^{3/2} \exp\left(-\frac{eE_{ion}}{kT_e}\right). \qquad (9)$$

Here, $g_i / g_a$ is the ratio of statistical weights of ground state and ionized state (for Argon this ratio is equal to 6, see Ref. 49; h is the Planck's constant; $E_{ion}$ is the ionization energy of argon atoms.

Because the arc is short (shorter than the radii of both electrodes) and currents are below 100 A (arc self-magnetic field is small and should not affect plasma profiles), flow in the inter-electrode region can be considered truly 1D; derivatives are predominantly along the arc axis. Using this assumption, we rewrite equations of conservation of fluxes for heavy and charged particles:

$$\Gamma_a + \Gamma_i = const = 0, \tag{10}$$

$$\Gamma_i - \Gamma_e = j/e = const. \tag{11}$$

Here, $j$ is the current density, which we use as the input parameter of the model. At the boundaries of the plasma domain (walls of the electrodes) the total flux of heavy particles is equal to zero, therefore it is zero everywhere (equation (10)).

**Energy transport equations:**

Energy transport of electrons and heavy particles gas is described by equations:

$$\nabla \cdot \left[ \frac{k}{e} T_e \left[ (2.5 + A_i^{(e)} + A_a^{(e)}) \vec{\Gamma}_e + \left( A_a^{(e)} \frac{n_e}{n_a} - A_i^{(e)} \frac{n_e}{n_i} \right) \vec{\Gamma}_i \right] \right] = \nabla \cdot (\lambda_e \nabla T_e) + \vec{j}_e \cdot \vec{E} - Q^{rad} - Q^{e-h} - Q^{ion}, \tag{12}$$

$$\nabla \cdot \left( \frac{k}{e} 2.5 T [\vec{\Gamma}_a + \vec{\Gamma}_i] \right) = \nabla \cdot (\lambda_h \nabla T) + Q^{e-h} + \vec{j}_i \cdot \vec{E}. \tag{13}$$

Here $\lambda_e = (\lambda_{e,i}^{-1} + \lambda_{e,a}^{-1})^{-1}$, $\lambda_h = \lambda_{h,i} + \lambda_{h,a}$ are the thermal conductivities of electron gas and heavy particles, $\lambda_{e,i} = 3.2 k n_e n D_{ei} / n_i$, definitions of $\lambda_{e,a}$, $\lambda_{h,i}$, $\lambda_{h,a}$ can be found in Ref. 13.

$Q^{e-h} = A^{e-H}(T_e - T)$ is the volumetric heat exchange between electrons and heavy particles,

$A^{e-H} = 8 n_e^2 \sigma_{ei} \sqrt{\frac{2 k m_e T_e}{\pi}} \frac{k}{m}$,

$Q^{ion} = s_i E_{ion}$ is the heat source due to ionization/recombination processes, and

$$Q^{rad} = 2.6 \times 10^{25} W/m^3 \frac{p}{1 atm} \left( \frac{1K}{T_e} \right)^{2.52} \exp\left( -\frac{1.69 \times 10^5 K}{T_e} \right) \tag{14}$$

is the volumetric heat loss due to radiation; it is supposed that plasma is optically thin and the radiation from plasma is a net sink of energy[13].

$A_i^{(e)} = \frac{1}{1+P} C_{tdei}$, $A_a^{(e)} = \frac{P}{1+P} C_{tdea}$

are kinetic coefficients[39].

Left-hand side of equations (12) and (13) represents the convective heat flux carried by electrons or heavy particles respectively. First terms on the right-hand sides of equations (12) and (13) correspond to the thermal conduction. The second terms on the right-hand sides of these equations represent effects of the Joule heating.

Note that due to differences in mobilities of electrons and ions, the ion flux is typically significantly smaller than the electron flux: about ten times smaller in the near-cathode region and about two orders of magnitude smaller elsewhere. Therefore, the second term under "nabla" operator on the left-hand side of Eq. (12) can be neglected.

Substitution of Eq. (10) into Eq. (13) makes the left-hand side of Eq. (13) equal to zero. Therefore all the heat flux of heavy particles is transported by thermal conductivity.

**The Poisson equation for the self-consistent electric field:**

$$e(n_e - n_i) = \varepsilon_0 \nabla \cdot \vec{E}. \tag{15}$$

Outside the space-charge sheaths, Gauss's law can be reduced to the quasi-neutrality approximation:

$$n_e = n_i. \tag{16}$$

**Equation of state:**

Plasma is treated as a mixture of ideal gases of electrons and heavy particles:

$$n_e k T_e + (n_a + n_i) k T = p. \tag{17}$$

Pressure variation is determined by Eq. (18). Substitution of Eq. (15) into Eq. (18) gives an algebraic relation for pressure:

$$p + \varepsilon_0 \frac{E^2}{2} = const = p_0, \tag{18}$$

where the second term in the left-hand side part is often referred as electrostatic pressure. $p_0$ is reference pressure or pressure in a far-away location where there is no electric field. It is an input parameter of the model (1 atm, in the considered case). Note that even in the cathodic space-charge sheath where electric field is at its highest, it is about 10 V per 1µm and the electrostatic pressure is about $10^3$ Pa, which is much lower than $p_0$. Therefore the total gas and plasma pressure in Eq. (17) can be approximated by a constant pressure, $p_0$ throughout the arc.

## II.2. Boundary conditions including sheath effects

To define the solution, boundary conditions are required at the plasma-electrode surfaces. The temperature of heavy particles should be equal to temperature of the electrode surface:

$$T = T_{electrode,tip}. \tag{19}$$

The temperature of the electrode can either be pre-defined (in case of water-cooled anode[50]) or has to be determined from solution of the heat balance equation as described in the next section.

Boundary conditions for other equations are formulated in terms of fluxes. Net electron flux at the plasma-electrode boundary is a composition of two oppositely directed fluxes: plasma electrons $\Gamma_{e,b}^{plasma}$ and electrons emitted from the electrode surface $\Gamma_{e,b}^{emiss}$:

$$j_e = e\left(\Gamma_{e,b}^{emiss} - \Gamma_{e,b}^{plasma}\right)e_{n,b}. \tag{20}$$

Here and further subscript *b* relates to the plasma-electrode boundary, $e_{n,b}$ denotes projection to the x-axis of the outward pointing unit normal from the electrode surface (see figure 1). Ion flux at the boundary is presented by ions fall from plasma onto the electrode surface:

$$j_i = e\,\Gamma_{i,b}^{plasma}e_{n,b}. \tag{21}$$

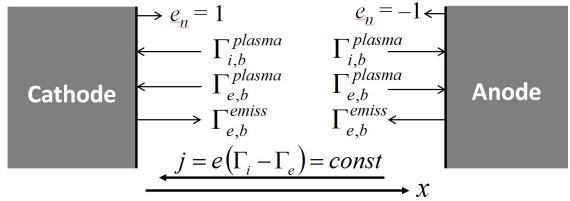

*Figure 1. Schematic of electric current composition at the surfaces of the electrodes.*

Addition of ion current density to the electron current density gives the total current density at the surface:

$$j = e\left(\Gamma_{e,b}^{emiss} + \Gamma_{i,b}^{plasma} - \Gamma_{e,b}^{plasma}\right)e_{n,b}. \tag{22}$$

Here, $j$ is the total current density, which is assumed constant and known in the 1D set-up. Relations for determining the boundary fluxes $\Gamma_{e,b}^{plasma}$, $\Gamma_{e,b}^{emiss}$, $\Gamma_{i,b}^{plasma}$ are given below by formulae (24) – (28).

Due to difference in masses and thermal velocities of electrons and ions, the electron thermal flux is often much higher than the current and the electron-flux-limiting sheaths form near the electrodes, where the electron and ion densities significantly deviate from the quasi-neutrality assumption. Charge separation results in the voltage drop in the sheath, which can be comparable to the total arc voltage. Sheath voltage drop suppresses or accelerates fluxes of charge particles in the sheath, depending on their charge and direction leading to positive or negative heat deposition in the sheath. Depending on direction of the electric current and its density, the wall temperature and plasma density at plasma-sheath boundary, the sheath voltage drop can be either positive or negative so that the balance of electric current (22) is satisfied.

We consider two approaches to modeling of the sheaths. In the first approach we resolve variations of the species densities and electric field in the sheath. In the second approach we use the fact that the sheath is typically much narrower than the inter-electrode gap, and can be assumed infinitely thin. Then effective sheath boundary conditions for fluxes of particles and heat can be applied at the plasma-sheath boundary without resolving the sheath.

**Effective boundary conditions for collisionless sheath:**

If the sheath is collisionless (its width is much narrower than the mean free paths of plasma particles) then it is convenient to use effective sheath boundary conditions[42,43]. Plasma can be modeled using quasi-neutrality assumption (16) instead of using Poisson's equation (15).

Relations below are given in a unified form applicable to both anode and cathode. The sheath voltage drop, $V_{sh}$, is considered positive if it gives a positive contribution to the total arc voltage. Therefore, the formulas should be used as is for the cathode, and sign of the sheath voltage should be reversed for the anode:

$$V_{sh,c} = V_{sh}, \ V_{sh,a} = -V_{sh}. \tag{23}$$

In case of positive sheath voltage drop $V_{sh}$, the ion current is determined by Bohm's condition:

$$\Gamma_{i,b}^{plasma} = n_e \sqrt{\frac{k(T + T_e)}{m_{Ar}}} . \tag{24}$$

For the negative sheath voltage, the ion current is:

$$\Gamma_{i,b}^{plasma} = \frac{1}{4} n_e \sqrt{\frac{8kT}{\pi m_{Ar}}} \exp\left(\frac{eV_{sh}}{kT}\right), \tag{25}$$

The electron current from plasma to the wall is:

$$\Gamma_{e,b}^{plasma} = \frac{1}{4} n_e \sqrt{\frac{8kT_e}{\pi m_e}} \exp\left(-\frac{e\max(V_{sh},0)}{kT_e}\right). \tag{26}$$

The electron emission current is:

$$\Gamma_{e,b}^{emiss} = \frac{j_R}{e} \exp\left(\frac{e\min(V_{sh},0)}{kT_{electrode}}\right), \tag{27}$$

where $j_R$ is the current density predicted by Richardson's emission law:

$$j_R = A_R T_{electrode}^2 \exp\left(-\frac{e(V_w + E_{Schott})}{kT_{electrode}}\right), \tag{28}$$

$A_R$ is the Richardson's constant, $V_w$ is the work function of the electrode material (4.5 V for tungsten), $E_{Schott}$ is the Schottky correction voltage (about 0.1 V, see Ref. 9 for instance).

Equations (24) – (28) for the fluxes of charged particles in combination with the balance relation for current density (22) allow determining the sheath voltage drop and it's sign.

The heat flux to the electron gas at the plasma-electrode boundary can be expressed as:

$$q_{e,b} = \Gamma_{e,b}^{emiss}(2.5kT_{electrode} + e\max(V_{sh},0)) - \Gamma_{e,b}^{plasma}(2.5kT_e + e\max(V_{sh},0)). \tag{29}$$

This flux is used as a boundary condition for the electron heat transport equation (12):

$$(2.5 + A_i^{(e)} + A_a^{(e)})(\Gamma_{e,b}^{emiss} - \Gamma_{e,b}^{plasma})kT_e - \lambda_e \frac{dT_e}{dx} e_{n,b} = q_{e,b}. \tag{30}$$

The heat flux to the electrode from the gas and the plasma can be expressed as:

$$q_{to\,electrode} = -\Gamma_{e,b}^{emiss}(2.5kT_{electrode} + e(V_w - \min(V_{sh},0))) +$$
$$+ \Gamma_{e,b}^{plasma}(2.5kT_e + e(V_w - \min(V_{sh},0))) + \Gamma_{i,b}^{plasma}\left(2.5\frac{k}{e}T + e(E_{ion} - V_w + V_{sh})\right) + \lambda_h \frac{dT}{dx} e_{n,b}. \tag{31}$$

**Solving for both plasma and sheath regions:**

If the sheath is strongly collisional (its width is much higher than the mean free paths of plasma particles) then it can be described by the fluid governing equations (1) – (17). The boundary conditions in this case are as follows. The heat flux to the electron gas is defined by:

$$q_{e,b} = 2.5kT_{electrode}\Gamma_{e,b}^{emiss} - 2.5kT_e \Gamma_{e,b}^{plasma}. \tag{32}$$

Zero number density condition can be used for the ion density (see Ref. 13):

$$n_i = 0. \tag{33}$$

Equation of current conservation (22) can in this case be used to determine the electron number density at the boundary:

$$j = e(\Gamma_{e,b}^{emiss} + \Gamma_{i,b}^{plasma} - \Gamma_{e,b}^{plasma})e_{n,b}$$

$$\frac{1}{4}n_e\sqrt{\frac{8kT_e}{\pi m_e}} = \Gamma_{e,b}^{plasma} = -j/(e_{n,b} e) + \Gamma_{i,b}^{plasma} + \Gamma_{e,b}^{emiss}. \tag{34}$$

## II.3. 1D Model of heat transfer in the electrodes

To account for cooling of electrodes by radiation from their sides, in quasi one-dimensional approximation we can neglect temperature variation in radial direction. Heat flux through the electrode is decreasing due to losses at its side surfaces and increasing due to Joule heating:

$$\pi r_{el}^2 \frac{d}{dx}\left(\lambda_{el}\frac{dT}{dx}\right) = 2\pi\left((T-T_{amb})\lambda_{gas} Nu + \varepsilon\sigma r_{el}(T^4 - T_{amb}^4)\right) + \pi r_{el}^2 j^2 \rho_{el}. \quad (35)$$

Here, $\lambda_{el}$ is the thermal conductivity of the electrode material (assumed to be constant, 170 W/m/K for tungsten), $r_{el}$ is the electrode radius, $T_{amb} = 300K$ is the ambient gas temperature, $\lambda_{gas}$ is the thermal conductivity of gas surrounding the electrode; the constant value of 0.1 W/m/K for $\lambda_{gas}$ was used, it corresponds to the gas temperature of about 3000 K (close to the temperature of the cathode tip where major heat losses take place). There is no need for more accurate resolution of the gas thermal conduction term, because it appeared to be of minor importance compared to the radiation term. $Nu$ is the Nusselt number taken to be equal to 1.1, see Ref. 51, $\sigma$ is the Stefan-Boltzmann constant, $\varepsilon$ is the emissivity of the electrode surface, taken to be equal to 1, $\rho_{el}$ is the electrical resistivity of the electrode material (assumed to be constant, small for metallic electrodes).

In the 1D model equation (35) is solved numerically with the heat flux coupled to heat flux from plasma at the arc side minus the surface radiation flux:

$$q_{el,tip} = \lambda_{el}\left(\frac{dT}{dx}\right)_{tip} = q_{to\,electrode} - q_{rad,tip}. \quad (36)$$

Here, $q_{to\,electrode}$ term is defined in Eq. (31), $q_{rad,tip}$ term takes into account radiation from front surface of the electrodes including mutual radiation[51]. Solution of Eq. (35) is used to define temperature of the electrode surface, which is used as a boundary condition in Eq. (19).

## III. Numerical procedure

As mentioned earlier, a set of governing equations is different depending on a sheath modeling approach utilized. In case of full sheath and plasma modeling approach, when the sheath and plasma regions are resolved with the same collisional model, one needs to take into account deviation between densities of electrons and ions, and treat them as two different variables and solve Poisson's equation (15) throughout the whole computational domain comprising of plasma and sheaths. In case of effective sheath boundary conditions, the sheath regions are not resolved by the plasma model and quasi-neutrality condition (16) can be utilized. Hence, one can reduce number of independent variables by exclusion of electron density from the governing equations. Two slightly different solution procedures were utilized depending on the sheath modeling approach.

In general case, equations (1) and (6) for ions can be transformed to one second order differential equation describing the ion transport. From equation (1) for ions, using relations between the fluxes of particles (10) and (11):

$$\vec{\Gamma}_i = \left(-kT\nabla n_i + n_i\left(k\nabla T + kC_i^{(e)}\nabla T_e + e\vec{E}\right) - \nu_{i,e}m_e\vec{j}/e\right)/\left(\nu_{i,a}m_{Ar}\right), \tag{37}$$

where $\nu_{k,j} = \dfrac{4}{3}\sqrt{\dfrac{8kT_{kj}}{\pi m_{kj}}}C_{kj}\sigma_{kj}n_j$ is the effective collision frequency of species $k$ with species $j$.

Substitution of ion flux (37) into equation (6) yields transport equation for the ions:

$$\nabla\left(n_i\vec{V}_i - D_i\nabla n_i\right) = s_i + \dfrac{m_e}{m_{Ar}}\dfrac{\vec{j}}{e}\nabla\left(\dfrac{\nu_{i,e}}{\nu_{i,a}}\right), \tag{38}$$

where $D_i = k(T+T_e)/(\nu_{i,a}m_{Ar})$, $\vec{V}_i = \left(k\nabla T + kC_i^{(e)}\nabla T_e + e\vec{E}\right)/(\nu_{i,a}m_{Ar})$.

The electric field can be determined from equation (1) for the electrons:

$$\vec{E} = -\dfrac{k}{e}\left(1+C_e^{(e)}\right)\nabla T_e - \dfrac{k}{e}T_e\dfrac{\nabla n_e}{n_e} + \dfrac{\vec{j}_e}{n_e e^2}m_e\left(\nu_{e,a}+\nu_{e,i}\right) + \dfrac{m_e}{e}\left(\nu_{e,a}\dfrac{\vec{\Gamma}_a}{n_a} + \nu_{e,i}\dfrac{\vec{\Gamma}_i}{n_i}\right). \tag{39}$$

In case of full plasma-sheath modeling, differential equations (38), (12) and (13) are solved with the boundary conditions defined in the previous section together with closures given by Eqs. (10), (11), (15), (17), (39) to form a fully closed system of governing equations. Due to non-linearity, differential equations (38), (12) and (13) were solved iteratively in the 1D code; and implicit scheme was used. At each iteration, the coefficients of these equations were computed using independent variables from the previous iteration, the source terms were linearized around values of independent variables from the previous iteration. The electric field can be obtained from Eq. (39), the electron density – from Eq. (15). Linearized equations are discretized using second order schemes resulting in tridiagonal matrices solved using tridiagonal matrix algorithm. For numerical stability implicit coupling between electric field and electron number density was used. Electron density in a numerator of the second term on the right-hand side of (39) was expressed via ion number density and electric field using Poisson's law (15). Therefore algebraic equation (39) for the electric field was transformed into the second order differential equation:

$$E - \dfrac{k\varepsilon_0}{e^2}\dfrac{T_e}{n_e}\nabla^2 E = -\dfrac{k}{e}\left(1+C_e^{(e)}\right)\nabla T_e - \dfrac{k}{e}T_e\dfrac{\nabla n_i}{n_e} + \dfrac{\vec{j}_e}{n_e e^2}m_e\left(\nu_{e,a}+\nu_{e,i}\right) + \dfrac{m_e}{e}\left(\nu_{e,a}\dfrac{\vec{\Gamma}_a}{n_a} + \nu_{e,i}\dfrac{\vec{\Gamma}_i}{n_i}\right). \tag{40}$$

The Neumann boundary conditions were used for this equation: gradient of electric field at the boundary is determined from a difference between ion number density (33) and electron number density (34) using Poisson's law (16).

When using the effective sheath boundary conditions, quasi-neutrality approximation for plasma is used ($n_e = n_i$) and the electric field can be excluded from the relation for the ion flux resulting in implicit and more numerically stable coupling of the variables. Substitution of Eq. (39) into Eq. (37) gives a relation for the ion flux (similar approach was used in Ref. 52):

$$\vec{\Gamma}_i = -D\nabla n_e - n_e \left(D_T \nabla \ln T + D_{Te} \nabla \ln T_e\right) - A_e \vec{\Gamma}_e, \tag{41}$$

where $D = \mu k (T + T_e)$ is the ambipolar diffusion coefficient,

$D_T = \mu k T$, $D_{Te} = \mu k \left(1 + C_e^{(e)} + C_i^{(e)}\right) T_e$ are the thermal diffusion coefficients,

$A_e = \mu \nu_{e,a} m_e$, $\mu = \left(0.5(\nu_{i,a} + \nu_{a,i}) m_{Ar} + \nu_{a,e} m_e\right)^{-1}$.

Substitution of the ion flux given by Eq. (41) into equation (6) and taking into account that $A_e \ll 1$ yields transport equation for ions:

$$\nabla \left(n_e \vec{V} - D \nabla n_e\right) = s_i + \vec{\Gamma}_e \cdot \nabla A_e, \tag{42}$$

where $\vec{V} = -D_T \nabla \ln T - D_{T_e} \nabla \ln T_e$.

A set of governing equations (42), (12) and (13) with closures (10), (11) and (39) for the electric field is solved iteratively in an implicit manner.

Note that in 1D formulation all derivatives are taken in primary coordinate along the axis of symmetry of the arc. Diffusion and convection in radial direction is assumed negligible. This assumption should be valid because the arc is short (distance between the electrodes is smaller than their diameters). Radial energy loss from plasma is taken into account by radiation term in equation (12). Radial energy losses from the side surfaces of electrodes are also taken into account and results show that for long and thin electrodes considered here these losses are of major importance.

## IV. Results and discussion

### IV.1. Benchmarking of 1D code and sheath models

For benchmarking of the code, 1D simulations of the near-cathode region of atmospheric pressure argon arc were performed for the same conditions as in Ref. 13. The current density was 5·10$^6$ A/m$^2$, fixed temperature 3500 K of the cathode surface was assumed. Full plasma-sheath modeling approach was employed; computational grid was strongly refined near the cathode to resolve variation of plasma parameters in the sheath. Comparison of our results with the simulations of Ref. 13 is given in figure 2; notice the logarithmic scale of x-axis. Very nice agreement on profiles of various plasma parameters was achieved – the profiles obtained in two simulations almost coincide. The regions of deviation from the ionization equilibrium, thermal equilibrium and quasi-neutrality are clearly visible and have different thickness. The sheath width appeared to be about one micron. This is less or comparable to the mean free paths of all plasma species, hence the sheath is weakly collisional and applicability of full plasma-sheath modeling approach is questionable in this case.

Benchmarking of two different sheath modeling approaches: i) full plasma-sheath fluid modeling; and ii) instead of sheath modeling use the effective boundary conditions for collisionless sheath (23) – (31) was also performed. Results of simulations of the near-cathode region with the effective boundary conditions are plotted in figure 2 with blue lines.

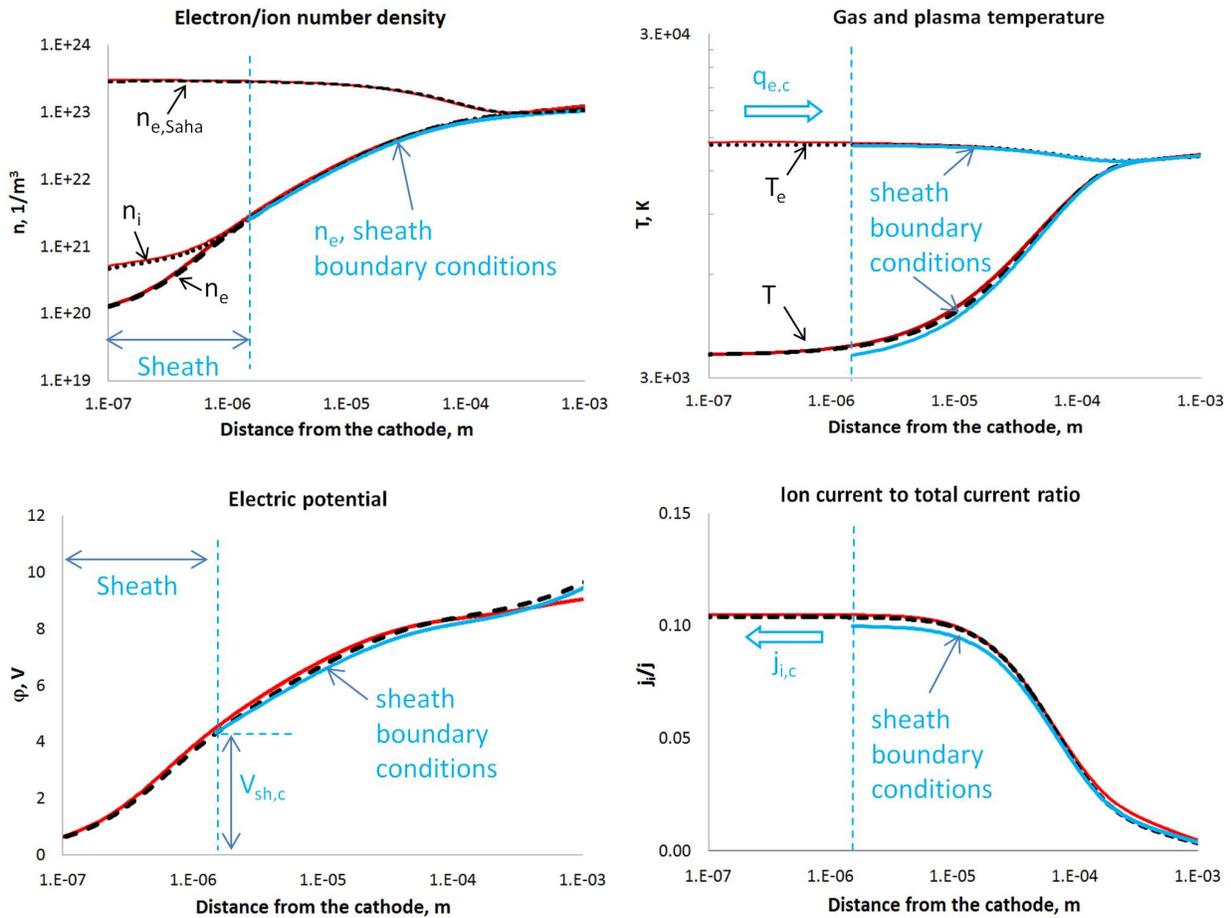

*Figure 2. Computational results for the near-cathode region. Red lines show results by N. Almeida[13]; black dashed/dotted lines – results of our 1D code with sheath resolution, blue lines – with the effective sheath boundary conditions.*

The results obtained with two sheath modeling approaches are in a very good agreement. This agreement is surprising at the first sight because the sheath models compared were supposed to be applicable to the opposite conditions: applicability of the full sheath-plasma model is justified in case of strictly collisional sheath (which is typical for pressures of tens of atmospheres), and the effective boundary conditions were designed for the collisionless sheath. Similarity of the heat fluxes to the cathode surface using these two approaches was already noted in Refs. 13 and 42. The new results show that not only the heat fluxes at the surface, but actually the whole profiles of plasma parameters appeared to be almost the same.

Comparison of performance of the sheath modeling approaches was also performed for the near-anode region. Pressure and current density were the same as for the near-cathode region. The anode temperature was 3900 K. The results are plotted in figure 3; notice that the anode is on the right side of the plots, logarithmic scale is used for the x-axis. Note that the electric potential profile appeared to be non-monotonic due to intensive electron emission from the anode surface: a negative electric field in the quasi-neutrality region and a positive electric field in the sheath region. One can define such near-electrode region as a double-layer. However, despite the complexity of the layer structure the results obtained with the two different approaches were in a very good agreement with each other.

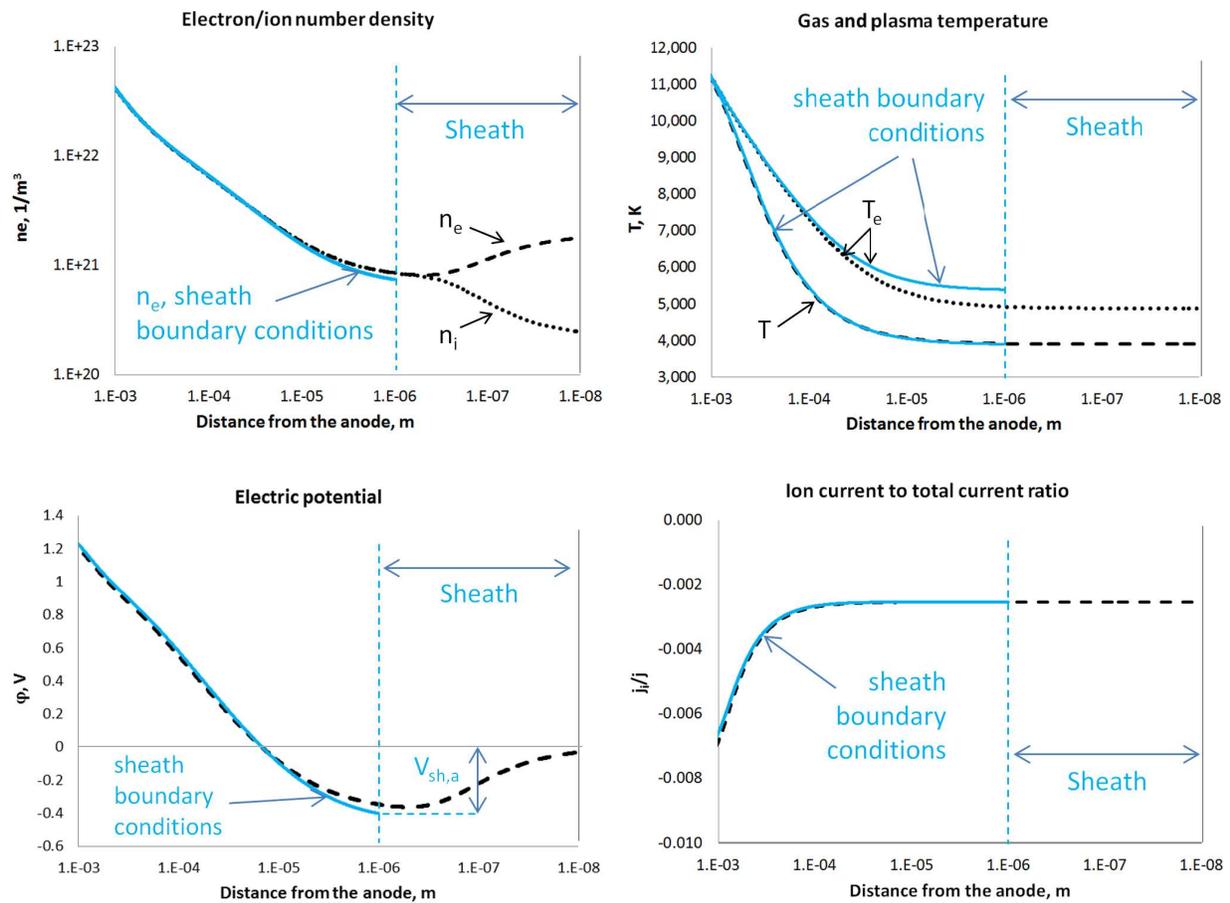

*Figure 3. Computational results for the near-anode region. Black dashed/dotted lines show results obtained with sheath resolution approach; blue lines - with the effective sheath boundary conditions.*

Independence of the results on the sheath modeling approach can be explained by the fact that both approaches capture the major effect determining voltage and current composition in the sheath and correspondingly plasma density at the sheath edge. This effect is reduction of fluxes of charged species by the Boltzmann exponent due to the sheath voltage. For the boundary conditions (23) – (31), this Boltzmann factor is implemented explicitly. In full modeling approach, the Boltzmann factor is manifested in exponential decrease of densities of charged particles with the electric potential profile according to Eq. (1) because the density gradients are the major effect in the sheath regions.

Note that the comparison of sheath modeling approaches was also performed in a wide range of current densities, $2 \cdot 10^6$ A/m$^2$ – $2 \cdot 10^7$ A/m$^2$ (typical for atmospheric pressure argon arc, see, for example, Refs.37, 53). In all the considered cases a good agreement was observed.

According to the results discussed above, both approaches were cross-verified against each other and can be claimed applicable for modeling of atmospheric pressure arcs.

## IV.2 Parametric studies of argon arc

### Discussion of the arc structure: arc column and near-electrode regions

Plasma density and temperature profiles for arcs of 2.5 mm and 5 mm lengths obtained in the simulations for various current densities are displayed in figures 4 and 5. Self-consistent heat transfer between the plasma and the electrodes 6 mm in diameter and 10 cm length was solved. Anodes of the arcs of different lengths are aligned with each other in the figures.

In the middle part of the long arc, ionization and thermal equilibriums take place: the plasma density profile is determined by the Saha equation (9), the temperatures of electrons and heavy particles are equal. Two sub-regions can be distinguished inside of the arc column: the complete equilibrium region where all plasma parameters are uniform (marked with green background in figures 4 and 5) and the local equilibrium region where plasma is non-uniform (white background in the figures 4 and 5). Near the electrodes the ionization and thermal equilibriums are not maintained; the non-equilibrium regions are marked with blue background in the figures.

As can be seen from figures 4 and 5, the near-electrode non-equilibrium regions are rather autonomic: profiles of plasma parameters are almost the same in the near-anode region of short and long arcs at the same current density. This is valid for all current densities considered, except for the lowest one ($2.5 \times 10^6$ A/m$^2$) at which the near-anode region tends to be longer than the short arc itself. But even in this case the profiles in the near-electrode non-equilibrium regions for short and long arcs are rather close. In the near-cathode region profiles of plasma parameters are close in cases of long and short arc for all current densities considered.

Note that autonomic behavior of the near-electrode regions would be expected in case of very long arc with an area of well-established plasma equilibrium between the near-electrode regions. However, in the presented results, absolute equilibrium with no variation of plasma parameters is established at most current densities. Therefore, plasma parameters in the arc column have rather small influence on the near-electrode regions.

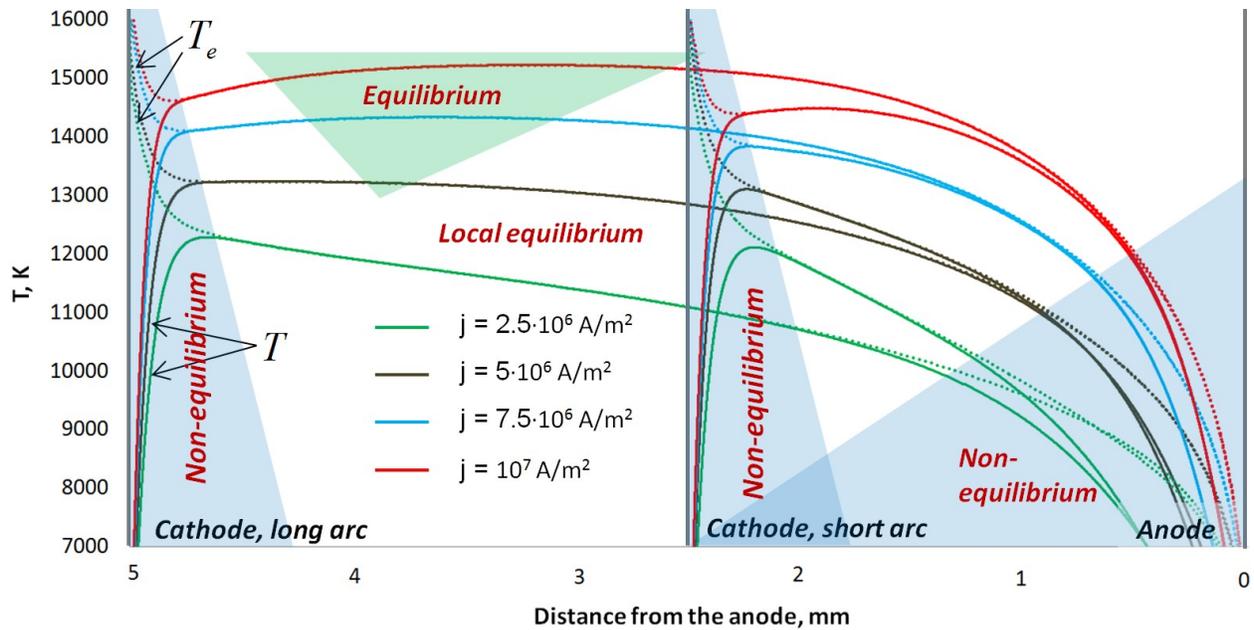

*Figure 4. Profiles of the gas temperature (solid lines) and electron temperature (dotted lines) of atmospheric pressure arcs 2.5 mm and 5 mm long at various current densities. Equilibrium and non-equilibrium regions are highlighted with different background colors. The non-equilibrium regions are up to several millimeters long and are comparable with the arc length. The near-anode region is typically longer than the near-cathode region and its length depends more strongly on the current density. Analytical relations for layers with dependence of current density can be found in the accompanying paper[44].*

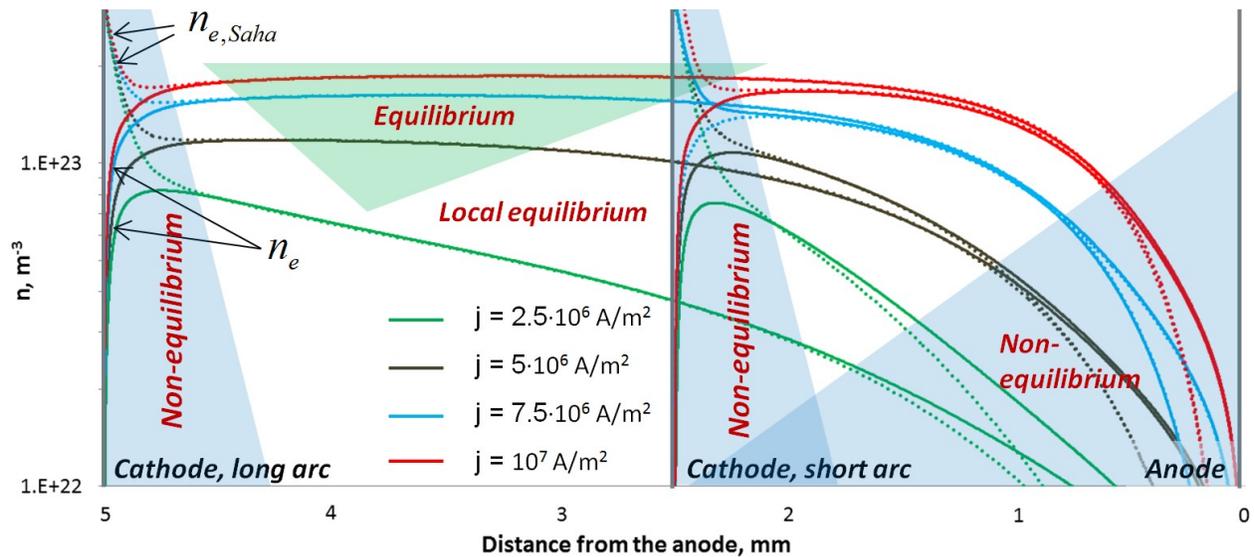

*Figure 5. Profiles of actual plasma density (solid lines) and equilibrium plasma density (dotted lines) obtained in 1D simulations at various current densities. Color notations are the same as in figure 4. Equilibrium and non-equilibrium regions are highlighted with different background colors.*

At higher current densities, the equilibrium region occupies significant part of the long arc, but is not observed in the short arc at any of the current densities considered. With decrease of the current density, the non-equilibrium and local equilibrium regions significantly elongate, especially near the anode, reaching several millimeters in width and resulting in vanishing of the equilibrium region in the 5mm-long arc.

This transformation results in qualitative change of the electric potential profile in the long arc (see figure 6). Plasma equilibrium region in the central part of the arc is characterized by constant electric field and linear growth of the electric potential (see figure 6). This region is commonly referred as a positive arc column. With increase of current density arc column becomes more pronounced: it elongates due to narrowing of the near-electrode regions, electric field in the column becomes stronger. Positive column cannot be distinguished at the lowest current density considered ($2.5 \cdot 10^6$ A/m$^2$).

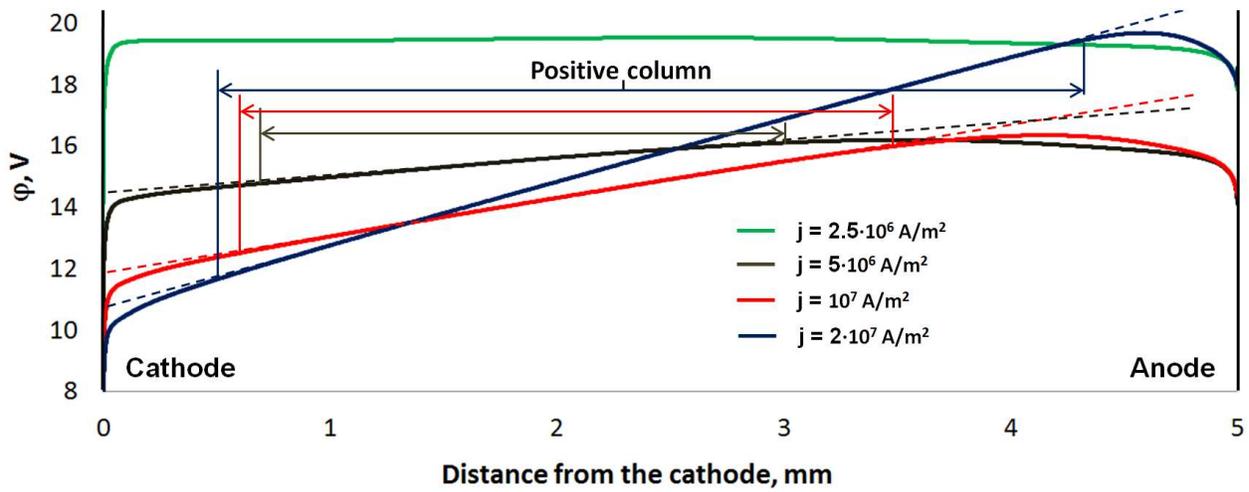

*Figure 6. Electric potential profiles obtained in 1D simulations for various current densities; zero potential is at the cathode surface. Linear profile of the electric potential corresponds to the "column" shown by the arrows. Its width increases with current density.*

In the near-cathode region electric potential significantly jumps from a reference zero value at the wall. Major part of the arc voltage is typically gained in this region; however, in case of higher current densities the role of the arc column increases and can become dominating at larger gap size. Near the anode, temperature and plasma concentration decrease causing electron diffusion towards the anode. It results in a decrease of electric field in order to preserve constant electron current density. In order to describe this variation of electric field it is convenient to express electron current density as a composition of electron mobility in the electric field, electron diffusion and thermal diffusion. This relation can be obtained from equation (39) neglecting the last term:

$$\vec{j}_e = \sigma \vec{E} + \sigma \frac{k}{e} T_e \nabla \ln n_e + \sigma \frac{k}{e}\left(1 + C_e^{(e)}\right) \nabla T_e, \qquad \sigma = \frac{n_e e^2}{m_e \left(\nu_{e,a} + \nu_{e,i}\right)} \qquad (43)$$

Here, $\sigma$ is the electrical conductivity (not to confuse with cross-sections).

Contributions of different electron current density components are displayed in figure 7 for the arc 2.5 mm length and current density $5\cdot10^6$ A/m$^2$. As mentioned earlier, no equilibrium region is present in the short arc. As seen from figure 7, the diffusion component is the largest throughout the arc, and the electric field does not play dominant role anywhere.

Near the electrodes, at distances about 0.5 mm, due to significant gradients of plasma density, the magnitude of the diffusion component becomes several times higher than total current density. It results in significant increase in the absolute value of the electric field in these areas in order to keep the total current constant. Near the cathode, the diffusion component is negative, the electric field is positive and vice versa near the anode. Therefore, significant heat source and heat sink are located in the near-cathode and the near-anode plasma, respectively.

The thermal diffusion plays significant role throughout the arc. It is comparable to other electric field components and should be taken into account in the accurate modeling; however, it is typically several times smaller than the diffusion component and does not lead to any qualitative effects. Note that the same conclusions are valid for longer arcs and other current densities not displayed in figure 7.

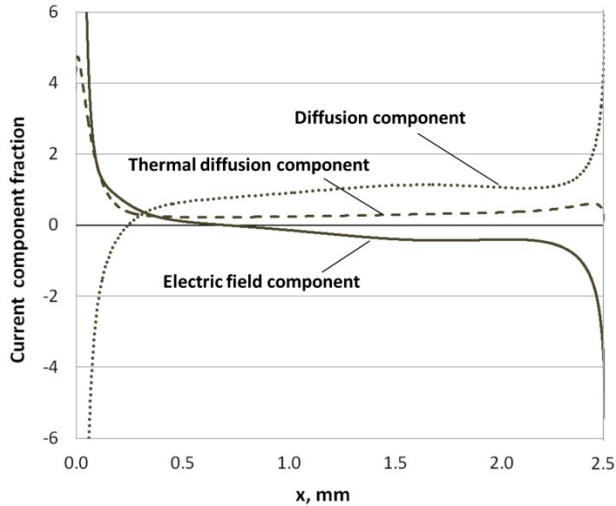

Figure 7. Profiles of electron current component fractions (43) along the arc for $j=5\cdot10^6$ A/m$^2$. Note that close to the electrodes diffusion component and field component are much higher than the current density and therefore almost balance each other. Thermal diffusion is crucial near the cathode only.

Note that the electron-ion collision frequency is much higher than the electron-atom one in most of the arc ($\nu_{e,i} \gg \nu_{e,a}$ in the denominator of parameter $\sigma$, equation (43)). Ratio of the collision frequencies, or parameter $P$ defined in equation (2), is about $(\sigma_{ea}n_a)/(\sigma_{ei}n_e)$ or $(\sigma_{ea}/\sigma_{ei})(1-\alpha_i)/\alpha_i$ where $\alpha_i = n_e/(n_e+n_a)$ is ionization degree. The ratio of cross-sections $\sigma_{ea}/\sigma_{ei}$ is about $0.01$ at temperatures below 18 000 K. According to figure 8, ionization degree is much higher than $0.01$ almost everywhere in the arc for all current densities considered. Hence, the parameter $P$ is small compared with unity throughout the arc, including major parts of the near-electrode regions. This statement will be used in the analytical arc model presented in the accompanying paper[44].

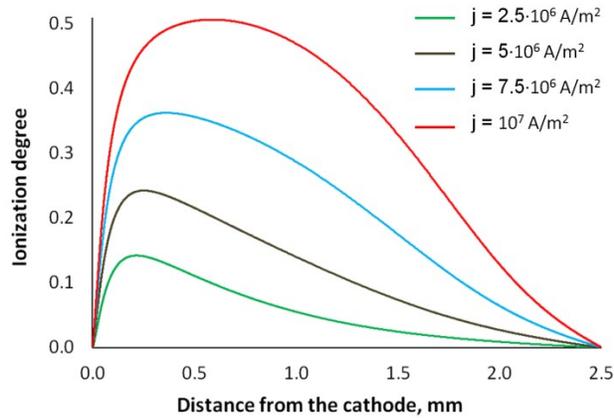

*Figure 8. Ionization degree profiles in the short (2.5mm) arc for various current densities. Ionization degree increases with current density.*

**Discussion of the near-cathode region**

Significant electric potential jump near the cathode (see figure 6) results in heating the electrons, as can be seen in figure 4. Temperature of heavy particles on the contrary decreases towards the cathode to reach temperature of the electrode at its surface. Provided the electron density is in ionization equilibrium ($n_{e,\,Saha}$, dotted lines in figure 5), increase in the electron temperature near the cathode would lead to increase of plasma density; whereas the actual plasma density decreases due to non-equilibrium fast acceleration of ions onto the cathode surface, as described by Bohm's condition (24) for positive sheath. Deviation between actual and equilibrium number densities exceeds an order of magnitude and results in high net production of ions that move towards the cathode due to electric field and density gradient. Ion current in the near-cathode region grows from zero values outside the region to about 15% to 25% of arc current at the cathode surface (see figure 9). Therefore, according to Eq. (22), electric current at the cathode surface is mostly carried by emitted electrons (75% - 85%; current of electrons from plasma is negligible because it is suppressed by voltage drop in the sheath, see Eq. (26)). This qualitative picture is in accordance with predictions of the previous theoretical models of the cathodic region[7,8], however quantitative disagreement in values of ion current ratio is observed with these studies. Opposite trend for ion current fraction was observed in simulations[13] where fixed cathode temperature was used. Therefore it is important to take heat transfer in the electrode into account in order to accurately describe the near-cathode plasma.

Electron emission requires certain temperature of the electrode surface, as specified by Eq. (28). The cathode temperature slightly goes up with increase of the total current density (see figure 10) but in general remains in the vicinity of 3500 K. Temperature of the electrode is determined by the energy balance between the electrode and the arc plasma, therefore heat transfer through the electrode should be taken into account, and interfacial conditions should be used.

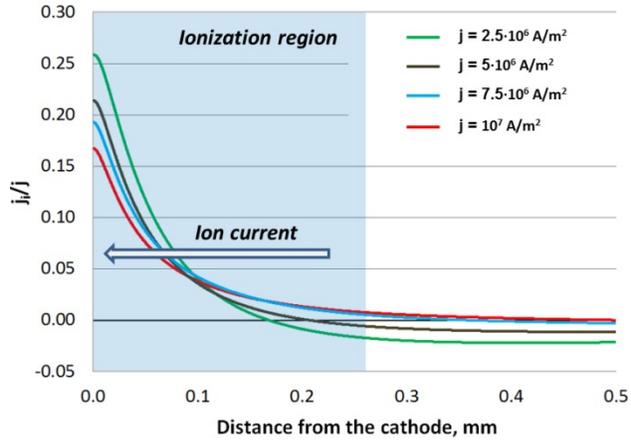

*Figure 9. Profiles of ion current fraction in the near-cathode region of atmospheric pressure arc for various current densities. Most of the generated ions move towards the cathode.*

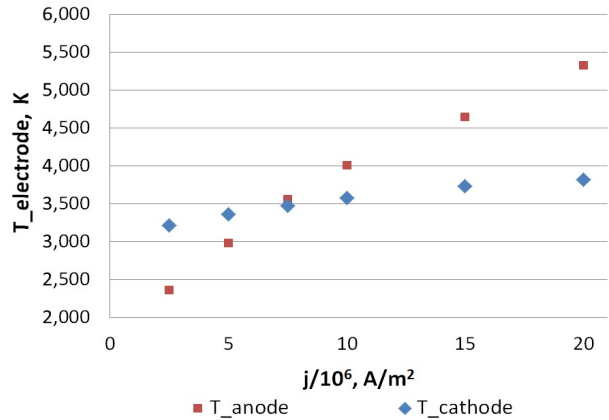

*Figure 10. Temperatures of the electrodes surfaces as functions of the current density. The cathode temperature doesn't change much because cathode should provide the necessary electron emission strongly depending on temperature. There is no such limitation at the anode.*

Heat fluxes in the plasma transferred by electrons and heavy particles are plotted in figures 11a and 11b correspondingly; positive values correspond to the heat fluxes directed away from the cathode (having positive projection to the x-axis). Conductive heat fluxes of the electrons and heavy particles correspond to the first terms on the right-hand sides of Eqs. (12) and (13) correspondingly. Electron convective heat flux corresponds to the left-hand side of Eq. (12). Ions carry their ionization energy; corresponding heat flux, $\Gamma_i E_i$, represents convective heat flux transferred by heavy particles. As mentioned earlier, heavy component as a whole does not move (see Eq. (10)), therefore its thermal energy does not contribute to the convective heat transfer.

As seen from figure 11a, electron convective heat flux is significant throughout the arc. Electrons gain significant thermal energy in the vicinity of the cathode and transfer it as they drift towards the anode.

Conductive heat transferred by the electrons is significant only at lower current densities and in the near-cathode region, where electron temperature gradients are high.

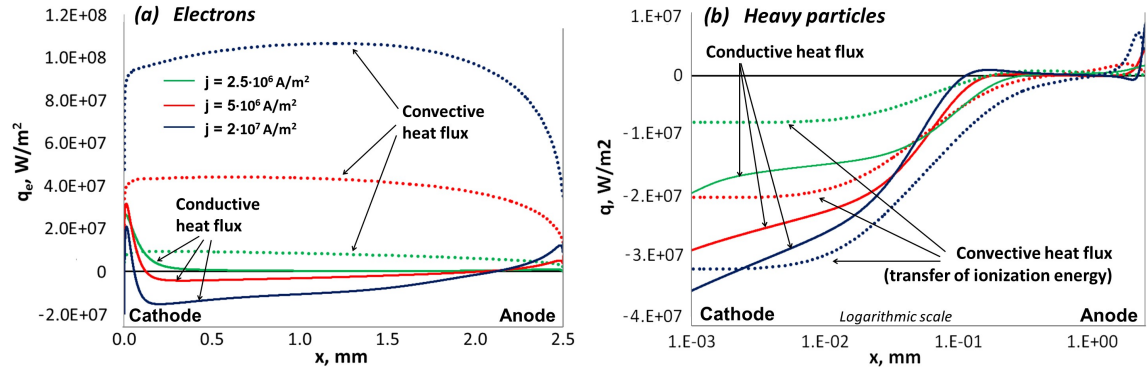

*Figure 11. Profiles of heat fluxes carried by electrons (a) and heavy particles (b). Positive values correspond to heat fluxes directed away from the cathode. Although in the plasma bulk thermal conduction is not significant, it plays important role close to the cathode.*

It should be noted that electrons gain significant amount of energy in the cathodic space-charge sheath: as can be seen from figure 12, about 2/3 of the near-cathode voltage drop is attributed to the sheath. Electrons emitted from the cathode surface accelerate significantly in the weakly collisional sheath and then exchange the energy in collisions with plasma species, mostly with plasma electrons rather than with heavy particles, due to difference in masses. High heat deposition in a very thin near-electrode region manifests in a sharp increase of the electron temperature at the cathode surface (see figure 4).

Generally speaking, there is a region in the vicinity of the cathode where the emitted electrons are not thermalized with plasma electrons. However, thickness of this region is about several electron mean free paths, i.e., several microns, much smaller than the length scale at which variation of the electron temperature occurs. Therefore, description of the near-cathode plasma with fluid model should be valid.

Hot plasma electrons in the near-cathode region transfer significant portion of their thermal energy to heavy particles in elastic and inelastic collisions corresponding to ionization and heating, see last two terms on the right-hand side of Eq. (12). Heavy particles bring this energy back to the cathode (see figure 11b) via thermal conduction and convection. These two constitutive parts of the heat flux are comparable in the near-cathode region and are both small outside the region; for illustrative purposes logarithmic scale was used for x-axis of figure 11b. As mentioned earlier, convective heat flux to the cathode is associated with ionization energy ions bring to the cathode during recombination at the surface.

In the previous theoretical studies[7,8,9] conductive heat flux to the cathode was neglected leading to overestimation of the ion current fraction. This question is addressed in more details in the accompanying paper[44].

The statement of low conductive heat transfer at the plasma boundary of the near-cathode region was previously used in theoretical study[9], but in that paper it was used as an assumption, without a proof.

Validity of this assumption was especially questionable for short arcs. In the present study we provide a justification based on the results of simulations of the whole arc. This simplification will be used in the arc model presented in the accompanying paper[44]. Quantitative asymptotic relations for the near-cathode region voltage drop, electron temperature, ion current and for size of the region is presented in the accompanying paper[44].

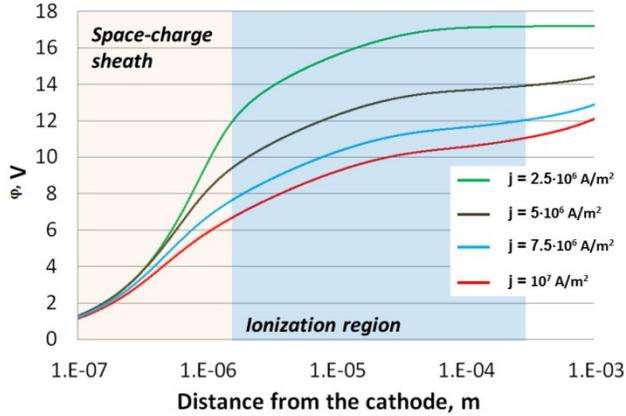

*Figure 12. Electric potential profiles in the near-cathode region of atmospheric pressure arc for various current densities. Voltage drop inside the cathode sheath constitutes a major fraction of the total voltage in the region. Note that x-axis scale is logarithmic.*

### Discussion of the near-anode region

The temperature of heavy particles decreases and reaches temperature of the anode at the electrode surface. The electron temperature is close to temperature of heavy particles due to collisional heat exchange (term $Q^{e-h}$ in equations (12) and (13)). Because the net electron flux is directed from plasma towards the anode, electron gas is losing energy at the boundary. It is easy to explain by rewriting equation (29) for the electron heat flux at the anode boundary in the following form:

$$q_{e,b} = -\Gamma_{e,b}\left(e\max(V_{sh},0) + 2.5kT_e\right) - \Gamma_{e,b}^{emiss}\left(2.5kT_e - 2.5kT_{electrode}\right), \tag{44}$$

where $\Gamma_{e,b} = \Gamma_{e,b}^{plasma} - \Gamma_{e,b}^{emiss}$ is net electron flux to the electrode. Apparently, first term in the right-hand side of Eq. (44) is negative. Second term is also negative if electron temperature at the boundary is higher than temperature of the electrode.

Due to negative heat flux to the electron gas, deviation between the temperatures of electrons and heavy particles is not strong, unlike in the near-cathode region. Electron temperature generally follows the downtrend of heavy particles temperature in the near-anode region (see figure 4), though does not reach temperature of the electrode; the simulations predicted electron temperature at the anode surface of about 5500 K, weakly depending on current density.

Electrons bring energy to the anode and heat it up. The energy flux brought to the anode by the electrons can be expressed from equation (31):

$$q_{\text{to anode by electrons}} = j_{e,a}V_w + j_{e,a}\max(V_{sh,a},0) + j_{e,a}2.5kT_e + 2.5k\Gamma_e^{emiss}(T_e - T_{electrode}). \qquad (45)$$

Here, $j_{e,a} = e\Gamma_{e,a}$ is the net electron current density to the anode. It is very close to total current density, because the ion current at the anode is small. The largest term in equation (45) is the first term on the right-hand side due to large value of work function compared to the temperature[27]. The convective heat flux is not high due to relatively low temperature; sheath voltage drop typically does not exceed 1 V, as will be shown below. Therefore heat flux to the anode is roughly proportional to current density, and as a result, temperature of the anode tip significantly increases with current density (see figure 10).

Low electron temperature near the anode results in significant decrease of equilibrium plasma density ($n_{e,\text{Saha}}$; see figure 5). Actual plasma density also decreases, though is higher than the equilibrium density (net recombination of ions takes place in the near-anode region). Due to significant decrease of plasma density, electron diffusion plays important role in near-anode region, driving electrons towards the anode. It results in strongly negative electric field (see figures 6, 7) and in negative voltage in the near-anode layer.

Electric potential profiles in the vicinity of the anode are plotted in figure 13 for two different current densities and for anodes with and without water cooling. In case of cooled anodes two temperatures of the anode surface were considered: 1 000 K and 3 000 K. For both temperatures, the electron emission from the anode surface (equation (28)) is negligible and, as can be seen from the figure, the potential decreases fast in the vicinity of the anode, and the sheath voltage drop is negative to suppress excess of electrons from plasma moving to the anode. Anode without additional cooling reaches temperatures above 4000 K at the current densities considered. Electron emission is very high at such conditions, and sheath voltage drop is positive in order to suppress the emission.

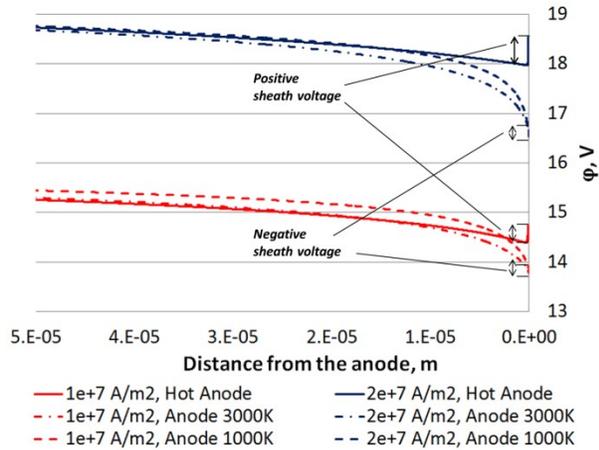

Figure 13. Electric potential profiles in the near-anode region of 5 mm arc; zero potential at the cathode.

In figure 14 the voltage drop in the anode sheath and whole near-anode non-equilibrium layer are plotted as functions of current density. Anodes with and without water cooling are considered. Because for cooled anode the voltages do not depend significantly on particular value of the anode temperature, the results are plotted for one temperature only (1000 K).

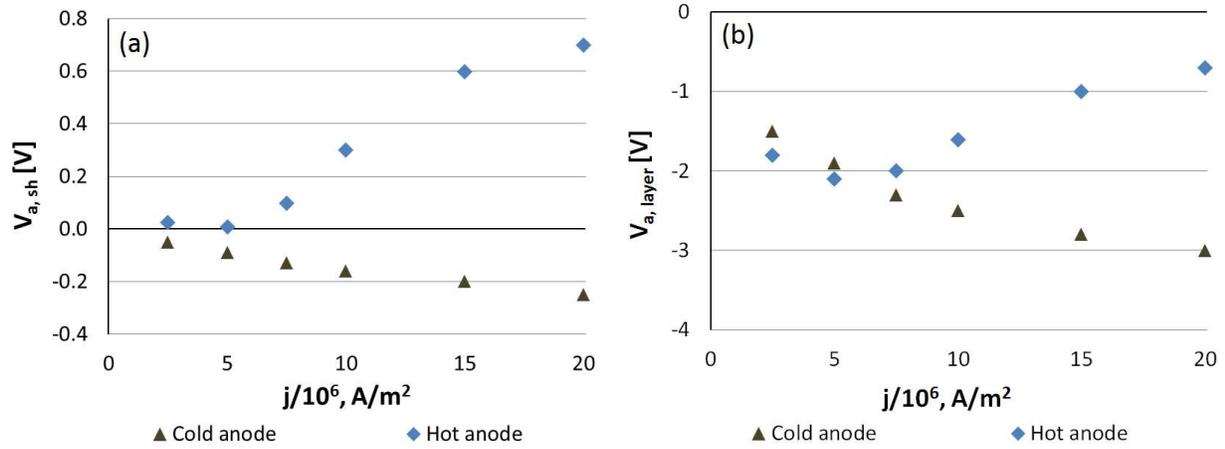

*Figure 14. Voltage drop in the near-anode sheath (a) and ionization non-equilibrium layer (b) as functions of current density. Note that the voltage drop in the anode sheath changes its sign from a negative (relatively cold anode, no electron emission) to a positive in the case of hot electron emitting anode.*

As can be seen from figure 14, in the case of cold anode voltage drop in the sheath is negative for all current densities considered. Voltages in the space-charge sheath and in the non-equilibrium layer as a whole decrease with current density (absolute values of the voltages increase). Whereas in case of hot anode, sheath voltage drop is positive and increases up to 0.7 V; voltage in the non-equilibrium layer increases only at low current densities but generally positive trend is observed, which is in a qualitative agreement with anode voltage measurements[54] in a carbon arc with hot anode. It leads to a conclusion that in case of cold anode arc constriction in the near-anode layer is energetically advantageous contrary to the hot anode case.

Accurate quantitative asymptotic relations for voltage and temperature in the non-equilibrium region and its size can be found in the second paper of the series[44].

## Significance of thermal and ionization non-equilibrium effects

Additional computational runs were performed to study effects of thermal and ionization non-equilibrium on integral arc characteristics and profiles of plasma parameters. Local thermal equilibrium (LTE) was assumed in these computational runs: electron temperature was taken equal to temperature of heavy particles. One energy balance equation was solved instead of equations (12) and (13):

$$\nabla \cdot \left( \frac{k}{e} T \left[ \left( 2.5 + A_i^{(e)} + A_a^{(e)} \right) \vec{\Gamma}_e + \left( A_a^{(e)} \frac{n_e}{n_a} - A_i^{(e)} \frac{n_e}{n_i} \right) \vec{\Gamma}_i \right] \right) = \nabla \cdot \left( (\lambda_e + \lambda_h) \nabla T \right) + \vec{j} \cdot \vec{E} - Q^{ion} - Q^{rad}. \quad (46)$$

Boundary condition (19) was used with a simple relation for heat flux to the electrode:

$$q_{to\,electrode} = j\left(V_{sh} - V_w - 2.5\frac{k}{e}T\right) + \Gamma_i^{plasma} e E_{ion} + (\lambda_e + \lambda_h)\frac{dT}{dx}e_n. \tag{47}$$

In addition to LTE assumption, the local ionization equilibrium (also often referred as chemical equilibrium, LCE) was assumed in one of the computational runs: Saha equation (9) was used to determine plasma density instead of the ion transport equation (42); ion production rate $s_i$ in the ionization heat term $Q^{ion}$ was calculated using equation (6). Effective collision-less sheath boundary conditions were used for quasi-neutral plasma equations.

In figure 15 results of non-equilibrium simulations for 2.5 mm arc at 5x10$^6$ A/m$^2$ current density (dark lines) are compared to the results obtained with LTE assumption (red lines) and both LTE+LCE assumptions (green lines).

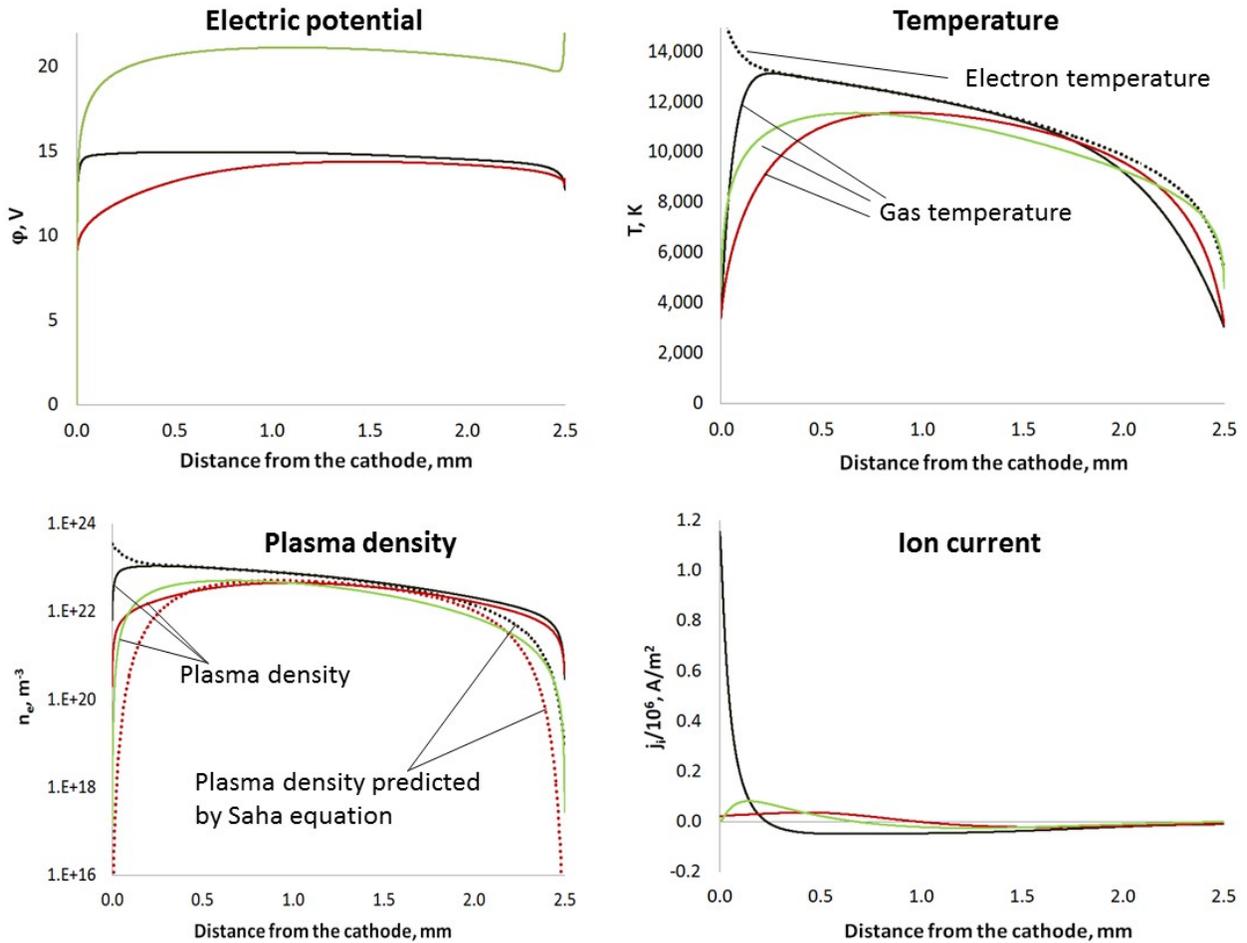

*Figure 15. Profiles of plasma parameters and electric potential in 2.5 mm arc at 5x10$^6$ A/m$^2$ current density obtained in 1D simulations under different assumptions: non-equilibrium simulations (black lines), LTE simulations (red lines) and LTE+LCE simulations (green lines).*

As can be seen from the figure, in the non-equilibrium case and the LTE case, the voltages in the near-cathode region and in the whole arc are almost the same as well as temperatures of the electrodes, but plasma profiles are significantly different, especially in the near-cathode region where significant deviation between temperatures of electrons and heavy particles should take place. The LTE simulations give lower electron temperature in the near-cathode region resulting in significantly less steep plasma density and electric potential profiles and prediction of drastically low value for the ion current.

In the "LTE+LCE" case, plasma density near the electrodes is strongly underestimated due to applying Saha equation at low temperatures. As a result, higher voltages in the near-electrode regions are required to heat up and maintain the plasma: about 5 V extra in each near-electrode region, i.e. 10 V in the arc total. Correspondingly, significantly higher temperatures of the electrodes are observed.

## IV.3. Validation against experimental data

Arc model was validated against experimental data of Refs. 40, 41 in which atmospheric pressure argon arc with cylindrical tungsten electrodes 3mm in diameter was run at arc currents of 30 A, 50 A and 100 A. Inter-electrode gap width was varied from 0.3 mm to 3.5 mm.

In figure 16 the arc voltage is plotted as a function of current density. Results of non-equilibrium 1D simulations with self-consistent heat transfer in the plasma and the electrodes are compared with the experimental data. Reasonable qualitative and quantitative agreement is observed. At larger inter-electrode gaps, the arc voltage linearly increases with the gap size. This behavior can be explained by elongation of equilibrium region of the arc column (see figures *4–6*). At smaller gaps (below 0.5–2mm, depending on the arc current), near-electrode non-equilibrium regions overlap and dependence of the arc voltage on current is different. Simulations show that the arc voltage decreases with current if the arc width is comparable with the non-uniform near-electrode layers. The same non-monotonic behavior is observed in the experimental data as well.

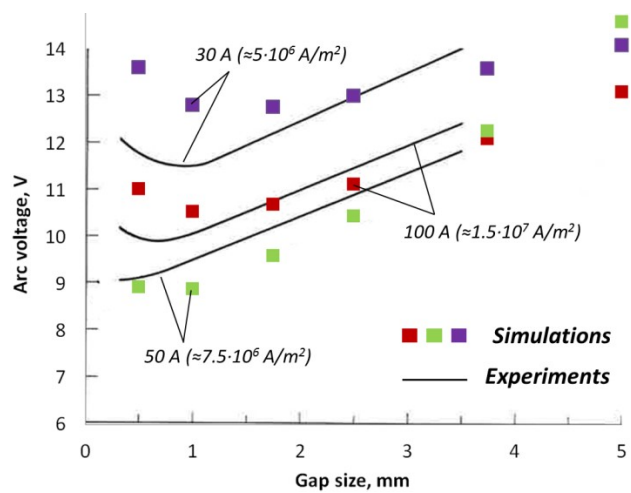

*Figure 16. Comparison of the computed arc voltage with the experiment data. The arc voltage as a function of the inter-electrode gap size for 3 currents. Experiment[40] – solid lines, 1D simulations – squares.*

# V    Conclusions

1D model of argon arc with cylindrical tungsten electrodes was implemented in a numerical code. The model features coupled heat transfer in the plasma and electrodes and accurate account of non-equilibrium effects in plasma including thermal, ionization non-equilibrium, electron diffusion, thermal diffusion and effects of space-charge sheaths. The model was benchmarked against previous simulations[13] and validated against experimental data[40]. Good agreement was obtained.

Role of non-equilibrium plasma effects in arc self-organization was studied. It was shown that *the electron diffusion significantly affects the electric field and leads to its reversal near the anode*. *Thermal diffusion of electrons also plays significant role*. However, its effect is smaller than that of classical diffusion.

Thermal non-equilibrium effect ($T_e \neq T_a$) is important, especially in the near-cathode region where significant deviation between temperatures of electrons and heavy particles take place, see figure 4. *Disregarding thermal non-equilibrium yields low electron temperature in the near-cathode region affecting plasma density and electric potential profiles and resulting in prediction of drastically low value for the ion current*, see figure 15. *Similarly, assuming thermal and ionization equilibriums results in prediction of drastically low value for the plasma density near the electrodes. Correspondingly, a higher voltage is required to maintain the arc plasma under the assumption of full thermal and ionization equilibriums*: additional 5 V in each near-electrode region, i.e., 10 V in the total arc voltage. Moreover, the temperatures of the electrodes are overestimated as well.

Two space-charge sheath modeling approaches were implemented in the code: (i) in the first approach: the sheath is assumed collisional and variations of the species densities in the sheath are resolved directly using the fluid model[13], and the Poisson equation is solved for the electric field; (ii) in the second approach: the sheath is assumed collisionless and effective sheath boundary conditions for fluxes of heat and charged particles[42,43] are applied at the plasma-sheath boundary without resolving the sheath, quasi-neutrality assumption is utilized for the plasma.

*Both collisional and collisionless space-charge sheath modeling approaches yielded the same plasma profiles for the near-electrode regions and entire arc*, see figures 2 and 3. Independence of the results on the sheath modeling approach can be explained by the fact that *both approaches capture the major effect determining voltage and current composition in the sheath and correspondingly plasma density at the sheath edge. This effect can be explained by the fact that electron density profile obeys the Boltzmann law due to high gradients of the electron density in anode and cathode regions*. For the collisionless boundary conditions (23)–(31), this Boltzmann factor is implemented explicitly. In addition ionization in the thin sheath regions does not contribute to the ion flux towards electrodes.

Parametric studies of short atmospheric pressure argon arc with tungsten electrodes were performed for various current densities and inter-electrode gap sizes, see figures 4 and 5. It was shown that *near-electrode regions can become up to several millimeters long significantly affecting arc characteristics*. The near-anode region is typically wider and its size more strongly depends on the current density

*Major part of the arc voltage typically falls in the near-cathode region; however, voltage drop in the arc column (where the Joule heating is locally balanced by radiation losses) increases with arc current and may become a considerable part of the arc voltage*, see figure 6. The voltage drop in the cathode sheath constitutes a major fraction of the voltage in the region in the near-cathode region. The cathode temperature is about 3500 K and does not change significantly with current. Significant part of the heat flux to the cathode is due to the thermal conduction of heavy particles.

Ion current fraction at the cathode is about 20% of the total current and decreases with increase of the total current density. This is in qualitative agreement with predictions of the previous models of the cathode region[7,8].

*Volt-Ampere characteristic (VAC) of the near-anode region depends on the anode cooling mechanism*, see figure 14. The anode voltage is negative. *In case of strong anode cooling (water-cooled anode) when the anode temperature is sufficiently low so that the thermionic emission from anode is negligible, the anode voltage decreases with current density (absolute value of the negative anode voltage drop increases)* due to stronger gradients of plasma density and temperature (see figures 4 and 5) at higher current densities. *Falling VAC of the near-anode region suggests the arc constriction near the anode*. *Without anode cooling*, the anode temperature increases significantly with current density, leading to drastic increase in the thermionic emission current from the anode. Correspondingly, *the potential barrier near the anode forms to limit the thermionic emission, and the total anode voltage increases with current density – the opposite trend in the VAC is observed as compared with the case of strong cooled anode*.

## Acknowledgements

The authors are grateful to Vlad Vekselman (PPPL, NJ), Yevgeny Raitses (PPPL, NJ), Mikhail Shneider (Princeton University, NJ), Nelson Almeida (Universidade da Madeira, Portugal), Mikhail Benilov (Universidade da Madeira), Ken Hara (Texas A&M University, TX) and Marina Lisnyak (Université d'Orléans, France) for fruitful discussions and valuable input.

The research is funded by the U.S. Department of Energy (DOE), Office of Fusion Energy Sciences.## References

[1] C. Journet, W.K. Maser, P. Bernier, A. Loiseau, M. L. delaChapelle, S. Lefrant, P. Deniard, R. Lee, J.E. Fischer, Nature **388**, 756 (1997).
[2] K. Ostrikov, A.B. Murphy, J. Phys. D: Appl. Phys. **40**, 2223 (2007).
[3] X.Q. Fang, A. Shashurin, G. Teel, M. Keidar, Carbon **107**, 273 (2016).
[4] S. Yatom, J. Bak, A. Khrabryi, Y. Raitses, Carbon **117**, 154 (2017).
[5] S. Yatom, R.S. Selinsky, B.E. Koel, Y. Raitses, Carbon **125**, 336 (2017).
[6] Y.W. Yeh, Y. Raitses, Bruce E. Koel, N. Yao, Scientific Reports **7**, 3075 (2017). doi:10.1038/s41598-017-03438-w
[7] W.L. Bade, J.M. Yos, AVCO Report RAD-TR-62-23 (1962).
[8] B.Ya. Moizhes, V.A. Nemchinskii, Sov. Phys. Tech. Phys. **17**, 793 (1972).
[9] M.S. Benilov, A. Marotta, J. Phys. D: Appl. Phys. **28**, 1869 (1995).
[10] R.M.S. Almeida, M.S. Benilov, G.V. Naidis, J. Phys. D: Appl. Phys. **33**, 960 (2000).28